\documentclass[conference]{IEEEtran}
\IEEEoverridecommandlockouts
\usepackage{cite}
\usepackage{amsmath,amssymb,amsfonts}
\usepackage{algorithmic}
\usepackage{graphicx}
\usepackage{textcomp}
\usepackage{xcolor}

\newcommand{\softplus}{\operatorname{softplus}}
\newcommand{\daggersup}{\textsuperscript{\textdagger}}

\def\BibTeX{{\rm B\kern-.05em{\sc i\kern-.025em b}\kern-.08em
    T\kern-.1667em\lower.7ex\hbox{E}\kern-.125emX}}
\begin{document}

\title{Generalizable Detection of Audio Deepfakes}

\author{\IEEEauthorblockN{Jose A. Lopez}
\IEEEauthorblockA{\textit{Intel Labs} \\
Arlington, MA, USA \\
jose.a.lopez@intel.com}
\and
\IEEEauthorblockN{Georg Stemmer}
\IEEEauthorblockA{\textit{Intel Labs} \\
Munich, BY, Germany \\
georg.stemmer@intel.com}
\and
\IEEEauthorblockN{Héctor Cordourier Maruri}
\IEEEauthorblockA{\textit{Intel Labs} \\
Guadalajara, JAL, Mexico \\
hector.a.cordourier.maruri@intel.com}
}

\maketitle

\begin{abstract}
In this paper, we present our comprehensive study aimed at enhancing the generalization capabilities of audio deepfake detection models. We investigate the performance of various pre-trained backbones, including Wav2Vec2, WavLM, and Whisper, across a diverse set of datasets, including those from the ASVspoof challenges and additional sources. Our experiments focus on the effects of different data augmentation strategies and loss functions on model performance. The results of our research demonstrate substantial enhancements in the generalization capabilities of audio deepfake detection models, surpassing the performance of the top-ranked single system in the ASVspoof 5 Challenge. This study contributes valuable insights into the optimization of audio models for more robust deepfake detection and facilitates future research in this critical area.
\end{abstract}


\section{Introduction}\label{sec:introduction}
The proliferation of tools for creating realistic deepfakes has led to a surge in their misuse by cybercriminals, posing a significant threat to individuals and society \cite{fbi_article_2024, delfino_2025}. Unfortunately, these tools can cause harm in both obvious and subtle ways. For instance, the case of the ``deepfake cheerleader mom'' highlights the potential for deepfake allegations to nearly result in wrongful criminal charges \cite{delfino_2025}. Moreover, deepfake allegations can be weaponized to restrict individuals' access to the justice system by driving litigation costs beyond their means \cite{delfino_2025}. Therefore, it is imperative for researchers to develop and democratize access to countermeasures.

Since 2015, the ASVspoof challenges have played a pivotal role in advancing research on countermeasures for automatic speaker verification \cite{asvspoof_paper_2015, asvspoof_data_2015, asvspoof_paper_2017, asvspoof_data_2017, asvspoof_paper_2019, asvspoof_data_2019, asvspoof_paper_2023, asvspoof_la_data_2021, asvspoof_pa_data_2021, asvspoof_df_data_2021, asvspoof_paper_2024, asvspoof_data_2024}. The resulting datasets and publications have provided invaluable resources for researchers, significantly contributing to our work in deepfake detection. Since 2022, the Audio Deep synthesis Detection (ADD) challenges have focused on specific issues not previously addressed by the ASVspoof competitions \cite{yi_add_2022}. These include the introduction of diverse background noises, hybrid real-fake audio samples, and the implementation of cutting-edge synthetic speech generation algorithms, all within the context of Mandarin language speech. In contrast, our work focuses on English language applications.

Bridging the gap between these advancements and the broader implications for deepfake technology, it is important to note that audio deepfakes are, in a sense, more accessible to create than their video counterparts. Indeed, the tools for their creation are widely available on open-source platforms. This accessibility stems from the fact that audio generators require less computational power to train, audio datasets are easier to collect, and audio-based solutions demand less storage and computational resources than video. However, these same factors also facilitate the development and training of countermeasures, enabling a more rapid defense against emerging threats. This fosters a dynamic environment of technological cat-and-mouse that bears a strong resemblance to the ongoing battle in antivirus detection.

Considering the trajectory of research in this cat-and-mouse game, the initial challenges saw researchers relying on smaller models such as ResNet, LCNN, and RawNet2, which we refer to as first-generation approaches \cite{muller_does_2022}. Although these models demonstrated strong performance, they tended to struggle with generalization, and to overfit to non-speech information like silence duration and high-frequency content \cite{chettri_2017, zhang_interspeech_2021, muller_silence_2021, muller_does_2022}. Subsequently, researchers discovered that pre-trained large language models (LLMs), like Wav2Vec2, trained on hundreds of thousands of hours of speech, offered significantly improved generalization performance \cite{wang_investigating_2022} and that these models tended to rely on the core speech frequency band between 0.1 kHz and 2.4 kHz \cite{wang_investigating_2022}. We refer to these as second-generation approaches, and we follow this direction in our work.

However, given the impracticality of deploying LLM ensembles on edge devices due to their speed and memory constraints, our research emphasizes and reports on the performance of individual models. In the forthcoming sections, we detail our contributions that build upon the foundational work of earlier countermeasure research, leading to what we believe is a state-of-the-art approach. In particular, our approach surpasses the equal-error-rate (EER) performance of the best reported single system of the ASVspoof 5 challenge.

In Section \ref{sec:data_sources}, we outline our data sources. Our model architecture is detailed in Section \ref{sec:architectures}, while Section \ref{sec:methodology} presents our training protocol. The findings from our exploration of pre-trained backbones are discussed in Section \ref{sec:backbone_exploration}. Section \ref{sec:loss_functions} introduces two novel loss function applications previously unexplored in deepfake detection literature. The data augmentations employed, along with the methodology used for ASVspoof5 data, are covered in Section \ref{sec:augmentation}. In Section \ref{sec:generalization}, we synthesize our learnings to present a model that has strong generalization across all test sets. Finally, Section \ref{sec:bias_and_reliability} addresses issues of bias and robustness, leading to our concluding remarks in Section \ref{sec:conclusions}.

\section{Data Sources}\label{sec:data_sources}
For training, we carried out experiments with training subsets from ASVspoof 2019 LA, ASVspoof 5, and our own collection that was produced using publicly available vocoders and speech from Speecon US. 

During evaluation, we utilized a range of datasets, including ASVspoof editions from 2015, 2019, 2021, and 5, as well as In-The-Wild (ITW) \cite{muller_itw_2022}, M-AILABS \cite{mailabs}, MLAAD v4 \cite{muller_mlaad_2024}, DeepFake Detection Challenge (DFDC) \cite{dfdc_kaggle}, and FakeAVCeleb \cite{favc_2021}. The selection of specific datasets for training as opposed to evaluation was guided by licensing restrictions. 

In our view, evaluating a classifier using datasets that include only one class is problematic, as performance gains may simply reflect a bias in the model's predictions, potentially at the expense of misclassifying the other class. Therefore, for M-AILABS and MLAAD v4 -- since M-AILABS is the source of authentic data for creating MLAAD v4, which contains only fake samples -- we generated a balanced sample set of 15,000 files, with approximately equal proportions from both M-AILABS and MLAAD v4. 

When evaluating the multi-modal DFDC dataset, we restricted our analysis to the training subset \cite{dfdc_kaggle}. The original annotations did not specify whether the audio or video components were fake. To address this, the community developed modality-specific annotations, as detailed in \cite{dfdc_annotations}, which we initially adopted for our experiments. However, further analysis using GCC-PHAT revealed minimal differences between real and fake audio categories. Consequently, we opted to reclassify this dataset as a source of authentic audio for our evaluation purposes.

\subsection{Proprietary Collection}\label{sec:proprietary_collection}
To expand our training data, we utilized the Speecon US dataset \cite{iskra_speecon_2002}, which features recordings from approximately 550 speakers. From each adult speaker, we selected four phonetically rich sentences and four spontaneous sentences, resulting in a total of 4,400 audio files. We subsequently filtered out files with less than four seconds of speech using voice activity detection \cite{silero_vad_2024}, and applied MP3 and M4A compression to increase diversity.

For the generation of the fake audio subset, we followed the approach outlined in \cite{wang_spoofed_training_2023}, using vocoders to create fake data. The underlying assumption is that using vocoders is equivalent to using an ideal text-to-speech engine or voice converter. To maximize diversity, we obtained 28 different publicly available vocoders from the Hugging Face platform \cite{huggingface}. These included models based on Generative Adversarial Networks (GANs) such as HifiGAN and MelGAN \cite{kong_hifigan_2020, kumar_melgan_2019}, signal processing methods like World \cite{masanori_world_2016}, flow-based techniques such as Waveglow \cite{prenger_waveglow_2018}, and neural source filters (NSFs) such as Hn-sinc-NSF \cite{wang_nsf_2020}. 

In the final step, we randomly selected 100,000 files, with approximately 90\% fake audio, to align with the proportions observed in the ASVspoof 2019 LA dataset.

\subsection{Data For Continual Training}
Some authors were able to obtain improvements by performing additional pre-training on the backbone model, otherwise known as continual training \cite{wang_can_large_scale_2024}. We conducted a few experiments with various speech datasets, including M-AILABS and LJ Speech \cite{ljspeech_2017}, but did not observe any benefit. For this additional pre-training, we used the Fairseq toolkit \cite{fairseq_2019}.

\subsection{Sampling Datasets}
Evaluating large datasets can be an extremely time-consuming process. For instance, the DFDC training set comprises over 117k files, while ASVspoof5 contains more than 1 million files, with its test subset alone accounting for 679k files. Utilizing samples can expedite evaluations and still provide a performance estimate that is sufficiently accurate. Consequently, we opted for a 1k-file sample from DFDC and, at times, a 100k-file sample from ASVspoof5. We will denote references to these samples with a superscript dagger, e.g., DFDC\daggersup.

\section{Model Architectures}\label{sec:architectures}
\begin{figure}[htbp]
\centering
\includegraphics[width=0.95\columnwidth]{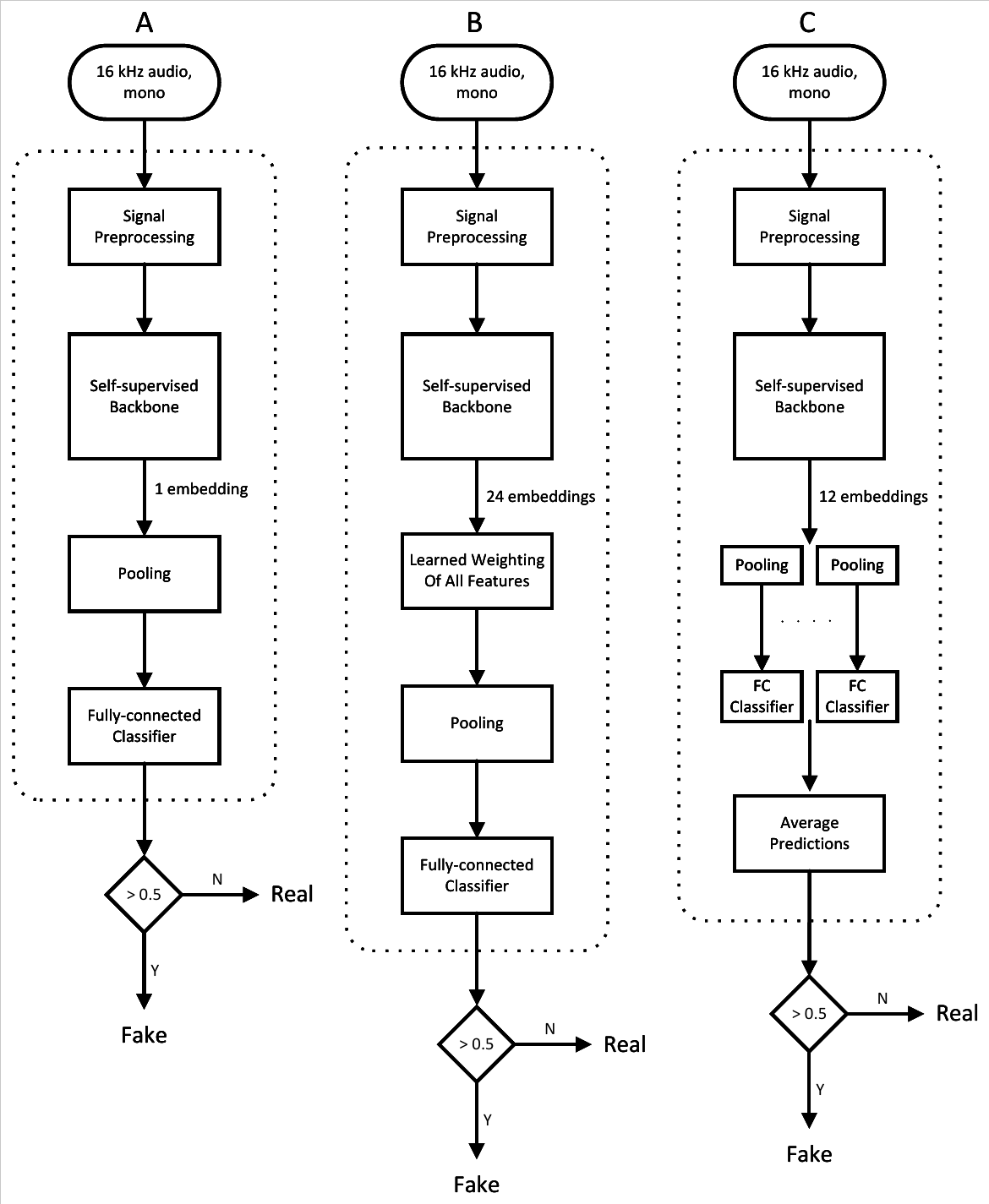}
\caption{Model Architectures}
\label{fig:architectures}
\end{figure}

During our research, we explored a range of architectures, some of which are depicted in Figure \ref{fig:architectures}. However, our findings aligned with those reported by Wang et al. \cite{wang_investigating_2022}, indicating that the use of more complex classifiers did not yield significantly different results. Consequently, in the interest of brevity and due to space constraints, we will only report the outcomes from experiments conducted with the simpler Architecture A.

In all the architectures we explored, the `signal processing' block includes signal standardization and bandpass filtering. This follows the filtering approach outlined by Tomilov et al. \cite{tomilov_2021} to mitigate the influence of spectral content outside the 0.3 kHz to 3.4 kHz band, for first-generation approaches. Additionally, in all experiments, we applied random scaling during training, adjusting the audio power to range from $1\times10^{-5}$ to 1.2. During validation and testing, we normalized the audio power to 1.0. The `pooling' block refers to temporal average pooling. Details regarding the fully connected classifier head are provided in Table \ref{tab:classifier_head}.

\begin{table}[htbp]
\caption{Classifier Head.}
\label{tab:classifier_head}
%\vskip 0.15in
\begin{center}
%\begin{small}
\begin{tabular}{|c|}
\hline
Linear(\text{embedding dim.}, 512) \\
\hline
LeakyReLU \\
\hline
Linear(512,64)\\
\hline
LeakyReLU \\
\hline
Linear(64,2)\\
\hline
\end{tabular}
%\end{small}
\end{center}
%\vskip -0.1in
\end{table}

\section{Experiment Methodology}\label{sec:methodology}
For all our experiments, we relied on the PyTorch library. We divided the model parameters into two groups: one containing the pre-trained backbone parameters and the other comprising the classifier head parameters. For the backbone, we applied no weight decay and set a learning rate of $1 \times 10^{-6}$. For the classifier head, we implemented a weight decay of $0.1$ and a learning rate of $1 \times 10^{-3}$. We used the AdamW optimizer in conjunction with a one-cycle learning rate schedule over 100 epochs. Lastly, during training, we used randomly selected 3.5-second audio segments.

For evaluation, we typically selected the model with the lowest validation loss and also used stochastic weight averaging (SWA)\cite{izmailov_2018}, as it often yielded better scores. We averaged 3 to 10 checkpoints post-training, enabling the selection of  neighboring checkpoints from the optimal segment along the validation loss curve.

\section{Backbone Exploration}\label{sec:backbone_exploration}
To assess the performance of various backbones, we trained each model minimizing the cross-entropy loss on the ASVspoof 2019 LA dataset, adhering to the splits specified in the dataset's metadata. For all models except Whisper, we used pre-trained weights available in the PyTorch audio library. For the Whisper backbone, we utilized the encoder from the medium-sized model provided by OpenAI, modified to process 3.5-second audio segments. This modification was straightforward, achieved by altering the parts of the code that enforced the default 30-second input duration.

As shown in Table \ref{tab:backbone_comparison}, the Wav2Vec2 models outperform the others. The larger, 1-billion-parameter version achieves the best average performance. However, due to its having three times as many parameters and being much slower to evaluate, we opted for the 300-million-parameter version for our experiments. It is therefore expected that a straightforward way to improve the metrics reported in this article would be to use the larger model, provided that the reader's computational constraints can accommodate it.

\begin{table}[htbp]
\caption{Performance of various backbones on test sets.}
\label{tab:backbone_comparison}
%\vskip 0.15in
\begin{center}
\begin{small}
\resizebox{\columnwidth}{!}{%
\begin{tabular}{|c|c|c|c|c|c|}
\hline
 & \textbf{Wav2Vec2} & \textbf{Wav2Vec2} & \textbf{Wav2Vec2} & \textbf{WavLM} & \textbf{Whisper}\\
 &\textbf{XLSR-53} & \textbf{XLS-R 300M} & \textbf{XLS-R 1B} & \textbf{Large} & \textbf{Medium}\\
\textbf{Test Set} & \textbf{(EER \%)} & \textbf{(EER \%)} & \textbf{(EER \%)} & \textbf{(EER \%)} & \textbf{(EER \%)}\\
\hline
FakeAVCeleb & 1.88 & 0.49 & \textbf{0.12} & 3.23 & 2.76 \\
\hline
ASVspoof2019\_LA\_test & 3.15 & 1.58 & 0.66 & \textbf{0.64} & 6.24 \\
\hline
ASVspoof2021\_LA\_progress & 6.69 & \textbf{2.98} & 3.46 & 4.20 & 8.12 \\
\hline
ASVspoof2021\_LA\_eval & 7.43 & \textbf{3.12} & 3.96 & 5.33 & 9.49 \\
\hline
ASVspoof2021\_LA\_hidden & 15.05 & \textbf{9.17} & 10.44 & 15.51 & 20.25 \\
\hline
ASVspoof2021\_DF\_progress & 4.68 & 1.94 & \textbf{1.46} & 8.36 & 7.35 \\
\hline
ASVspoof2021\_DF\_eval & 3.42 & 2.45 & \textbf{2.27} & 7.59 & 7.91 \\
\hline
ASVspoof2021\_DF\_hidden & 12.28 & \textbf{7.06} & 8.68 & 11.89 & 19.45 \\
\hline
ASVspoof2015 & 0.20 & 0.43 & \textbf{0.19} & 0.73 & 6.16 \\
\hline
In-The-Wild & 16.64 & 13.42 & \textbf{4.78} & 15.14 & 25.70 \\
\hline
\textbf{Average} & 7.14 & 4.26 & \textbf{3.60} & 7.26 & 11.34 \\
\hline
\end{tabular}
}
\end{small}
\end{center}
%\vskip -0.1in
\end{table}

\subsection{Score Aggregation}
To obtain the scores in Table \ref{tab:backbone_comparison}, we processed the entire test files. In subsequent experiments, we windowed the audio into segments of the same 3.5-second duration that was used during training, with a 0.5-second step size. This approach more closely aligned with the performance we can expect during deployment and slightly improved the EER. We also evaluated an overlap-and-average approach, and this resulted in marginally better results; however, as this would increase evaluation time, we decided it was not crucial for answering our research questions.

\section{Loss Functions}\label{sec:loss_functions}
Loss functions are a key part of any training procedure, so we explored various loss functions beyond the typical cross-entropy. Given that there are clearly easier and more difficult examples of fake audio, it was natural to incorporate focal loss to diminish the impact of the easier samples \cite{lin_focal_2018}. We believe the application of focal loss to this problem is novel. We also investigated techniques to reduce intra-class embedding distances, such as one-class softmax and center loss objectives\cite{wen_2016, zhang_oc_softmax_2021}. Center loss was also used in \cite{tomilov_2021}. However, one drawback of these techniques is that, upon convergence, the intra-class loss can compete with the class loss. We prioritize reducing class loss, so to address this issue, we modified the center loss: by incorporating a hinge, we can prevent this competition after reaching an adequate level, thus favoring the classification component. Without this modification, we found that adding center loss did not improve performance.

Focal loss is given by Equation \ref{eq:focal_loss}, where $p$ is the predicted probability and $\gamma$ is the tunable focusing parameter. We remind the reader that when $\gamma=0$, the loss reduces to cross-entropy. The majority of our experiments fixed $\gamma=2.0$. 

\begin{equation}
L_f(p_t) = -(1-p_t)^\gamma \log(p_t)
\label{eq:focal_loss}
\end{equation}

Center loss is given by Equation \ref{eq:center_loss}, where $N$ is the number of samples, $x_i$ is the embedding of the $i$th prediction, $c_{y_i}$ is the center associated with the class of the $i$th sample's target class. 

\begin{equation}
L_{\text{center}}(x,y) = \frac{1}{2}\sum_{i=1}^N \| x_i - c_{y_i} \|_2^2
\label{eq:center_loss}
\end{equation}

Our hinged modification is given by Equation \ref{eq:hinged_center_loss}, and the smooth version for implementation is given by Equation \ref{eq:smooth_hinged_center_loss}, where we fix $\beta=20.0$ in our experiments.
\begin{equation}
L_{\text{hinged}}(x,y) = \max(0, L_{\text{center}} - 1.0)
\label{eq:hinged_center_loss}
\end{equation}

\begin{equation}
L_{\text{smooth}}(x,y) = \softplus (\beta ( L_{\text{center}} - 1.0))
\label{eq:smooth_hinged_center_loss}
\end{equation}

One-class softmax is given by Equation \ref{eq:oc_softmax_loss}, where $N$, $x$, and $y$ are as above, $w_0$ is a learnable parameter, and $m_{y_i}$ are margin scalars, and $\alpha$ is a scale factor used in the equation. We use the implementation provided by the authors \cite{zhang_oc_softmax_2021}.

\begin{equation}
L_{\text{oc}}(x,y) = \frac{1}{N}\sum_{i=1}^N \log\left( 1 + \exp^{\alpha (m_{y_i}-w_0 x_i)(-1)^{y_i} } \right)
\label{eq:oc_softmax_loss}
\end{equation}

Table \ref{tab:loss_comparison} shows the comparison of experiments conducted using the XLS-R 300M backbone that was continually trained using M-AILABS and LJ Speech for a few epochs. From these and other experiments, we concluded that continual training may not improve performance. Additionally, we found that focal and hinged-center losses yielded better results than cross-entropy with one-class softmax losses. For the remainder of our experiments, we used focal and hinged center losses for training. 

\begin{table}[htbp]
\caption{Performance using various loss functions.}
\label{tab:loss_comparison}
%\vskip 0.15in
\begin{center}
\begin{small}
\resizebox{\columnwidth}{!}{%
\begin{tabular}{|c|c|c|c|}
\hline
 & \textbf{Cross-entropy \&} & \textbf{Cross-entropy \&} & \textbf{Focal \&}\\
 &\textbf{One-class Softmax} & \textbf{Hinged-center} & \textbf{Hinged-center}\\
\textbf{Test Set} & \textbf{(EER \%)} & \textbf{(EER \%)} & \textbf{(EER \%)}\\
\hline
FakeAVCeleb & \textbf{0.19} & 0.39 & 0.21 \\
\hline
ASVspoof2019\_LA\_test & 4.07 & 3.59 & \textbf{3.02} \\
\hline
ASVspoof2021\_LA\_progress & 5.92 & 5.08 & \textbf{4.24} \\
\hline
ASVspoof2021\_LA\_eval & 7.88 & 7.52 & \textbf{7.06} \\
\hline
ASVspoof2021\_LA\_hidden & 12.40 & \textbf{12.05} & 12.21 \\
\hline
ASVspoof2021\_DF\_progress & 5.89 & 5.29 & \textbf{4.14} \\
\hline
ASVspoof2021\_DF\_eval & \textbf{2.68} & 3.10 & 3.31 \\
\hline
ASVspoof2021\_DF\_hidden & \textbf{9.55} & 11.27 & 10.30 \\
\hline
ASVspoof2015 & \textbf{0.12} & 0.36 & 0.16 \\
\hline
In-The-Wild & 19.42 & \textbf{15.05} & 19.12 \\
\hline
M-AILABS\_MLAAD & 30.68 & 16.69 & \textbf{16.23} \\
\hline
ASVspoof5\_train\daggersup & 2.17 & 2.10 & \textbf{1.46} \\
\hline
ASVspoof5\_val\daggersup & \textbf{0.29} & 0.37 & 0.35 \\
\hline
ASVspoof5\_test\daggersup & 11.03 & 10.87 & \textbf{10.81} \\
\hline
\textbf{Average} & 9.15 & 7.61 & \textbf{7.57} \\
\hline
\end{tabular}
}
\end{small}
\end{center}
%\vskip -0.1in
\end{table}

\section{Data Augmentation Strategies}\label{sec:augmentation}
Data augmentation is a proven strategy in past ASVspoof challenges \cite{liu_towards_2023}. In this article, we explored several augmentation techniques, including simple additive white Gaussian noise (AWGN), RawBoost \cite{tak_rawboost_2022}, vocoded audio, and room impulse response (RIR) augmentation \cite{jeub_aachen_2009}. Table \ref{tab:configuration_differences} shows the configurations used for the initial round of data augmentation experiments. For AWGN, we mixed noise into the audio at signal-to-noise ratios (SNRs) ranging from 5 to 30 dB during training with a 50 percent probability ensuring that the original audio was still utilized during training. This approach was taken to avoid the situation where a model performs worse on clean data. For RawBoost, we employed the best-reported algorithm and implementation from \cite{tak_rawboost_2022}, and applied the augmentation 75 percent of the time. The vocoded Speecon data is described in Section \ref{sec:proprietary_collection}.

Table \ref{tab:augmentation_comparison} presents the results of the experiments, which all used ASVspoof2019 LA training data. Clearly, even the addition of AWGN can lead to performance gains; however, the configurations employing RawBoost were the most effective. Notably, all models achieved impressive results on FakeAVCeleb, ASVspoof2015, and, unexpectedly, the training and validation subsets of ASVspoof5. Although configuration D had the lowest average EER, we chose configuration E for further exploration at that time. This decision was based on its performance with In-The-Wild data and the fact that we had not included data from M-AILABS, MLAAD v4, and ASVspoof 5 in our evaluations.

\begin{table}[htbp]
\caption{Configurations for data augmentation.}
\label{tab:configuration_differences}
%\vskip 0.15in
\begin{center}
%\begin{small}
\begin{tabular}{|c|c|}
\hline
\textbf{Configuration} &  \textbf{Data Augmentation}\\
\hline
A &  none\\
\hline
B &  AWGN\\
\hline
C & AWGN, proprietary\\
\hline
D & RawBoost\\
\hline
E & RawBoost, proprietary\\
\hline
\end{tabular}
%\end{small}
\end{center}
%\vskip -0.1in
\end{table}

\begin{table}[htbp]
\caption{Performance using various augmentation configurations.}
\label{tab:augmentation_comparison}
%\vskip 0.15in
\begin{center}
\begin{small}
\resizebox{\columnwidth}{!}{%
\begin{tabular}{|c|c|c|c|c|c|}
\hline
 &\textbf{Config. A} & \textbf{Config. B} & \textbf{Config. C} & \textbf{Config. D} & \textbf{Config. E} \\
\textbf{Test Set} & \textbf{(EER \%)} & \textbf{(EER \%)} & \textbf{(EER \%)} & \textbf{(EER \%)} & \textbf{(EER \%)} \\
\hline
FakeAVCeleb & 0.33 & 0.91 & 0.13 & 0.47 & \textbf{0.12} \\
\hline
ASVspoof2019\_LA\_test & 0.95 & 0.58 & 1.55 & 0.31 & \textbf{0.28} \\
\hline
ASVspoof2021\_LA\_progress & 2.55 & 2.75 & 5.30 & \textbf{1.85} & 2.49 \\
\hline
ASVspoof2021\_LA\_eval & 3.42 & 3.37 & 5.56 & \textbf{2.46} & 3.72 \\
\hline
ASVspoof2021\_LA\_hidden & 10.92 & 9.75 & 8.93 & 9.18 & \textbf{8.06} \\
\hline
ASVspoof2021\_DF\_progress & 1.70 & 1.27 & 1.73 & \textbf{0.50} & 0.61 \\
\hline
ASVspoof2021\_DF\_eval & 2.23 & 1.88 & 1.51 & 1.84 & \textbf{0.92} \\
\hline
ASVspoof2021\_DF\_hidden & 9.19 & 7.27 & 6.97 & 7.13 & \textbf{6.09} \\
\hline
ASVspoof2015 & 0.12 & 0.10 & \textbf{0.08} & 0.18 & 0.14 \\
\hline
In-The-Wild & 5.78 & 6.57 & 4.44 & 4.29 & \textbf{2.12} \\
\hline
M-AILABS\_MLAAD & 13.43 & \textbf{12.89} & 15.85 & 13.10 & 13.33 \\
\hline
ASVspoof5\_train\daggersup & 0.97 & 0.92 & 1.56 & 0.48 & \textbf{0.39} \\
\hline
ASVspoof5\_val\daggersup & 0.57 & 0.54 & 0.25 & 0.61 & \textbf{0.17} \\
\hline
ASVspoof5\_test\daggersup & 10.90 & \textbf{10.65} & 18.35 & 11.87 & 19.26 \\
\hline
\textbf{Average} & 4.50 & 4.25 & 5.16 & \textbf{3.88} & 4.12 \\
\hline
\textbf{Average (FAVC-ITW)} & 3.72 & 3.45 & 3.62 & 2.82 & \textbf{2.46} \\
\hline
\end{tabular}
}
\end{small}
\end{center}
%\vskip -0.1in
\end{table}

\subsection{ASVspoof5}\label{ssec:asvspoof_5}
Upon further examination of why our models performed much worse on the test subset of ASVspoof5, we found that the degradation was not due to any particular attack; indeed, the configuration E model attained near-perfect accuracy on all attacks. Instead, the degradation was attributed to poor performance on the authentic audio of the test set, the only subset that underwent codec compression, which in some cases was extreme. We noted that the methods used in the top solution \cite{chen_asvspoof5_2024} were not very different from ours, except that they included codecs, resampling, and calibration. 

To gauge baseline performance, we initially trained a model using only ASVspoof5 data without augmentations. We discovered that the EER on the test set was halved simply by using the same source data. When we added Encodec, the most challenging codec in terms of performance on the data \cite{defossez_encodec_2022}, the resulting model performed similarly, indicating that the codec augmentation had little effect. However, adding other augmentations did have an impact. As shown in Table \ref{tab:asvspoof5_configurations}, incorporating AWGN, RIR, and RawBoost, with and without resampling (which involves resampling the audio to 8 kHz and back to 16 kHz), resulted in an EER that surpassed the best reported single system in the challenge, which attained an EER of 5.56, and was competitive with the top ensemble systems. In hingsight, we believe that resampling was somewhat redundant, as the bandpass filtering technique from \cite{tomilov_2021} had already been included.

\begin{table}[htbp]
\caption{Configurations and results for ASVspoof 5 experiments.}
\label{tab:asvspoof5_configurations}
%\vskip 0.15in
\begin{center}
\begin{small}
\resizebox{\columnwidth}{!}{%
\begin{tabular}{|c|c|c|}
\hline
 &   & \textbf{ASVspoof5 Test}\\
\textbf{Config.} &  \textbf{Data Augmentation} & \textbf{(EER \%)}\\
\hline
F &  none & 7.68\\
\hline
G & Encodec augmentation & 7.58\\
\hline
H & AWGN, RIR, RawBoost & \textbf{3.57}\\
\hline
I & AWGN, RIR, RawBoost, resampling & 4.98\\
\hline
\end{tabular}
}
\end{small}
\end{center}
%\vskip -0.1in
\end{table}

\section{Leveraging Learnings for Generalization}\label{sec:generalization}
To maximize generalization, we anticipated that using all of our training data (ASVspoof2019 LA, ASVspoof5, proprietary) would yield the best performance. However, in configurations J and K, shown in Table \ref{tab:generalization_configurations}, we observed that while the additional ASVspoof5 training data improved performance on some test sets, such as the hidden subsets of ASVspoof2021 and M-AILABS/MLAAD, it led to degradations on others, particularly In-The-Wild and ASVspoof5. The latter was unexpected, given that the ASVspoof5-only experiments, detailed in Table \ref{tab:asvspoof5_configurations}, indicated much lower EERs. The experiments described in configurations L-R aimed to restore performance on these test sets and achieve more balanced results across all our test sets. We discovered that the best performance was achieved using configuration R by omitting ASVspoof5 data from the training set and instead introducing its information through a teacher model trained solely on ASVspoof5. Table \ref{tab:data_and_distillation_comparisons} presents the results of these experiments, with averages computed using ASVspoof5\daggersup for comparison.

\begin{table}[htbp]
\caption{Configurations For Generalization Experiments.}
\label{tab:generalization_configurations}
%\vskip 0.15in
\begin{center}
\begin{small}
\resizebox{\columnwidth}{!}{%
\begin{tabular}{|c|c|c|c|}
\hline
\textbf{Config.} & \textbf{Training Data} & \textbf{Teacher} & \textbf{Data Augmentation} \\
\hline
J & ASVspoof2019\_LA, ASVspoof5\daggersup, proprietary & none & AWGN, RIR, RawBoost\\
\hline
K & ASVspoof2019\_LA, ASVspoof5\daggersup, proprietary & none & AWGN, RIR, RawBoost, \textbf{resampling}\\
\hline
L & ASVspoof2019\_LA, \textbf{ASVspoof5}, proprietary & none & AWGN, RIR, RawBoost, resampling\\
\hline
M & ASVspoof2019\_LA, \textbf{ASVspoof5\daggersup}, proprietary & \textbf{config. K} & AWGN, RIR, RawBoost, resampling\\
\hline
N & ASVspoof2019\_LA, ASVspoof5\daggersup, proprietary & \textbf{config. E} & AWGN, RIR, RawBoost, resampling\\
\hline
O & ASVspoof2019\_LA, ASVspoof5\daggersup, proprietary & \textbf{config. I} & AWGN, RIR, RawBoost, resampling\\
\hline
P & ASVspoof2019\_LA, proprietary & config I. & AWGN, RIR, RawBoost, resampling\\
\hline
Q & ASVspoof2019\_LA, proprietary & \textbf{config. H} & AWGN, RIR, RawBoost, resampling\\
\hline
R & ASVspoof2019\_LA, proprietary & config. H & AWGN, RIR, RawBoost\\
\hline
\end{tabular}
}
\end{small}
\end{center}
%\vskip -0.1in
\end{table}

\begin{table*}[htbp]
\caption{Results Of Experiments J-R.}
\label{tab:data_and_distillation_comparisons}
%\vskip 0.15in
\begin{center}
%\begin{small}
%\resizebox{\columnwidth}{!}{
\begin{tabular}{|c|c|c|c|c|c|c|c|c|c|}
\hline
 &\textbf{Config. J} & \textbf{Config. K} & \textbf{Config. L} & \textbf{Config. M} & \textbf{Config. N} & \textbf{Config. O} & \textbf{Config. P} & \textbf{Config. Q} & \textbf{Config. R}\\
\textbf{Test Set} & \textbf{(EER \%)} & \textbf{(EER \%)} & \textbf{(EER \%)} & \textbf{(EER \%)} & \textbf{(EER \%)}  & \textbf{(EER \%)} & \textbf{(EER \%)} & \textbf{(EER \%)} & \textbf{(EER \%)}\\
\hline
FakeAVCeleb & \textbf{0.13} & \textbf{0.13} & 0.15 & \textbf{0.13} & 0.14 & 0.14 & 0.14 & 0.14 & 0.14\\
\hline
ASVspoof2019\_LA\_test & 0.46 & 0.39 & 0.45 & 0.32 & \textbf{0.25} & 0.57 & 0.33 & 0.33 & 0.43\\
\hline
ASVspoof2021\_LA\_progress & 1.85 & 1.55 & 1.61 & 1.42 & 1.97 & 1.63 & \textbf{1.36} & 1.74 & 1.43\\
\hline
ASVspoof2021\_LA\_eval & 3.75 & 2.19 & 3.00 & 2.13 & 3.19 & 1.75 & \textbf{1.70} & 2.31 & 2.12\\
\hline
ASVspoof2021\_LA\_hidden & 6.30 & 5.57 & 5.65 & \textbf{5.56} & 7.04 & 7.91 & 8.60 & 8.98 & 7.82\\
\hline
ASVspoof2021\_DF\_progress & 0.66 & 0.38 & 0.48 & 0.34 & 0.34 & 0.59 & \textbf{0.33} & 0.41 & 0.44\\
\hline
ASVspoof2021\_DF\_eval & \textbf{0.63} & 0.75 & 1.01 & 0.77 & 0.81 & 1.71 & 2.13 & 2.21 & 1.71\\
\hline
ASVspoof2021\_DF\_hidden & 3.88 & 3.89 & \textbf{3.55} & 3.90 & 4.91 & 5.66 & 5.55 & 6.26 & 5.06\\
\hline
ASVspoof2015 & \textbf{0.04} & 0.07 & 0.07 & 0.07 & 0.14 & 0.35 & 0.13 & 0.15 & 0.17\\
\hline
In-The-Wild & 3.33 & 6.85 & 8.52 & 6.72 & \textbf{2.47} & 5.88 & 3.96 & 3.62 & 3.19\\
\hline
M-AILABS\_MLAAD & 5.75 & 4.81 & 5.05 & 4.81 & 6.44 & 6.72 & 4.72 & \textbf{3.84} & 4.42\\
\hline
ASVspoof5\_test\daggersup & 13.33 & 11.40 & 10.17 & 11.20 & 17.13 & 5.32 & 4.62 & \textbf{3.60} & 4.34\\
\hline
ASVspoof5\_test & - & - & - & - & - & 5.38 & 4.70 & \textbf{3.69} & 4.48\\
\hline
\textbf{Average} & 3.34 & 3.17 & 3.31 & 3.11 & 3.74 & 3.19 & 2.80 & 2.80 & \textbf{2.61} \\
\hline
\end{tabular}
%}
%\end{small}
\end{center}
%\vskip -0.1in
\end{table*}

\subsection{Calibrated Predictions}\label{ssec:calibrated_predictions}
Like Chen et al. \cite{chen_asvspoof5_2024}, we found that calibrating model predictions is generally beneficial, though not in every case. For this purpose, we employed Platt calibration \cite{platt_1999}, as described in Equation \ref{eq:platt_calibration}, where $p_t$ is the predicted fake score and $\hat{p}_t$ is the calibrated score. This procedure involves fitting the coefficients $a_i$ using a calibration dataset. Notably, the calibration function is monotonic, which preserves the order among the scores and does not impact threshold-independent metrics such as the EER and the Area Under the Curve (AUC). The primary benefit is the improvement of threshold-dependent metrics, including accuracy and the F1-score. In \cite{chen_asvspoof5_2024}, it is our understanding that the authors used an expanded version of Equation \ref{eq:platt_calibration} that incorporates additional information, such as speech quality and duration, by introducing extra coefficients. Although this increases expressivity, it no longer preserves the score ordering and, in our experiments, degrades generalization.

The results presented in Table \ref{tab:best_model_results} correspond to configuration R and were obtained using uncalibrated predictions with a threshold of 0.5, as calibration did not yield an improvement in performance. However, at other times, we found that using small samples, approximately 1,000 files, from the hidden subsets of ASVspoof2021 LA proved useful for calibration. Table \ref{tab:best_model_results}  also includes reference scores from the literature for comparison. Note that the reference scores were collected from numerous published experiments across several references, whereas our results are from a single model.

\begin{equation}
\hat{p}_t = \frac{1}{1+\exp\left(a_0 + a_1 p_t\right)}
\label{eq:platt_calibration}
\end{equation}

\begin{table*}[htbp]
\caption{Performance Of Model With Lowest Average EER.}
\label{tab:best_model_results}
%\vskip 0.15in
\begin{center}
%\begin{small}
\begin{tabular}{|c|c|c|c|c|c|}
\hline
\textbf{Test Set} &\textbf{F1-score} & \textbf{EER \%} & \textbf{Accuracy \%} & \textbf{AUC} & \textbf{Reference EER \%}\\
\hline
FakeAVCeleb & 0.9960 & 0.14 & 99.60  & 0.9998  & - \\
\hline
ASVspoof2019\_LA\_test & 0.9650 & 0.43  & 98.63  & 0.9998   & \textbf{0.13} \cite{wang_can_large_scale_2024} \\
\hline
ASVspoof2021\_LA\_progress & 0.9637  & \textbf{1.43} & 98.61  & 0.9982   & 4.06 \cite{wang_investigating_2022} \\
\hline
ASVspoof2021\_LA\_eval & 0.9618 & 2.12 & 98.58  & 0.9975  & \textbf{1.32} \cite{asvspoof_paper_2023} \\
\hline
ASVspoof2021\_LA\_hidden & 0.8024 & \textbf{7.82}  & 93.70  & 0.9782  & 9.53 \cite{wang_can_large_scale_2024}  \\
\hline
ASVspoof2021\_DF\_progress & 0.9669 & \textbf{0.44} & 98.77  & 0.9999  & 0.88 \cite{wang_investigating_2022}\\
\hline
ASVspoof2021\_DF\_eval & 0.9732 & \textbf{1.71}  & 99.70  & 0.9989  & 3.31 \cite{wang_can_large_scale_2024} \\
\hline
ASVspoof2021\_DF\_hidden & 0.8460 & \textbf{5.06} & 95.13  & 0.9876  & 6.11 \cite{wang_spoofed_training_2023}\\
\hline
ASVspoof2015 & 0.9814 & 0.17 & 99.54  & 0.9999  & \textbf{0.16} \cite{wang_investigating_2022}\\
\hline
In-The-Wild & 0.9004 & \textbf{3.19} & 90.29  & 0.9952  & 4.25 \cite{wang_can_large_scale_2024}\\
\hline
M-AILABS\_MLAAD & 0.8930 & 4.42 & 89.40  & 0.9887  & -\\
\hline
ASVspoof5\_test & 0.9230 & \textbf{4.48} & 95.42  & 0.9921  & 5.56 \cite{asvspoof_paper_2024} \\
\hline
DFDC & - & -  &  92.40   &  -  & - \\
\hline
\textbf{Average} & 0.9310 & 2.62 & 96.14  & 0.9947  & \\
\hline
\end{tabular}
%\end{small}
\end{center}
%\vskip -0.1in
\end{table*}

%DFDC* 91.90

\section{Assessing Bias And Reliability}\label{sec:bias_and_reliability}
Evaluations using the FB ASR Fairness dataset \cite{veliche_2024}, which contains annotations on age, gender, ethnicity, geographic location, and first language, did not indicate any kind of bias, correctly identifying samples as authentic across all categories. The few incorrect predictions corresponded to files with significantly lower speech quality.

\begin{figure}[htbp]
\centerline{\includegraphics[width=\columnwidth]{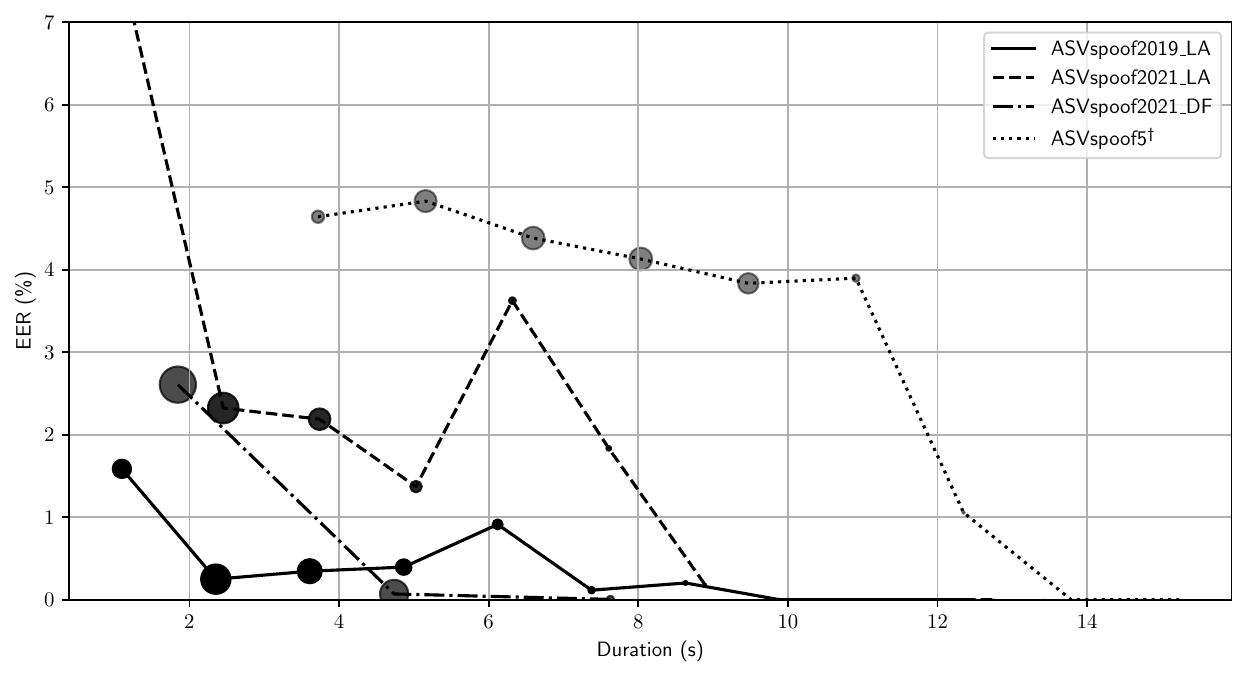}}
\caption{EER Distribution Across Duration Bins}
\label{fig:duration_bins}
\end{figure}

\begin{figure}[htbp]
\centerline{\includegraphics[width=\columnwidth]{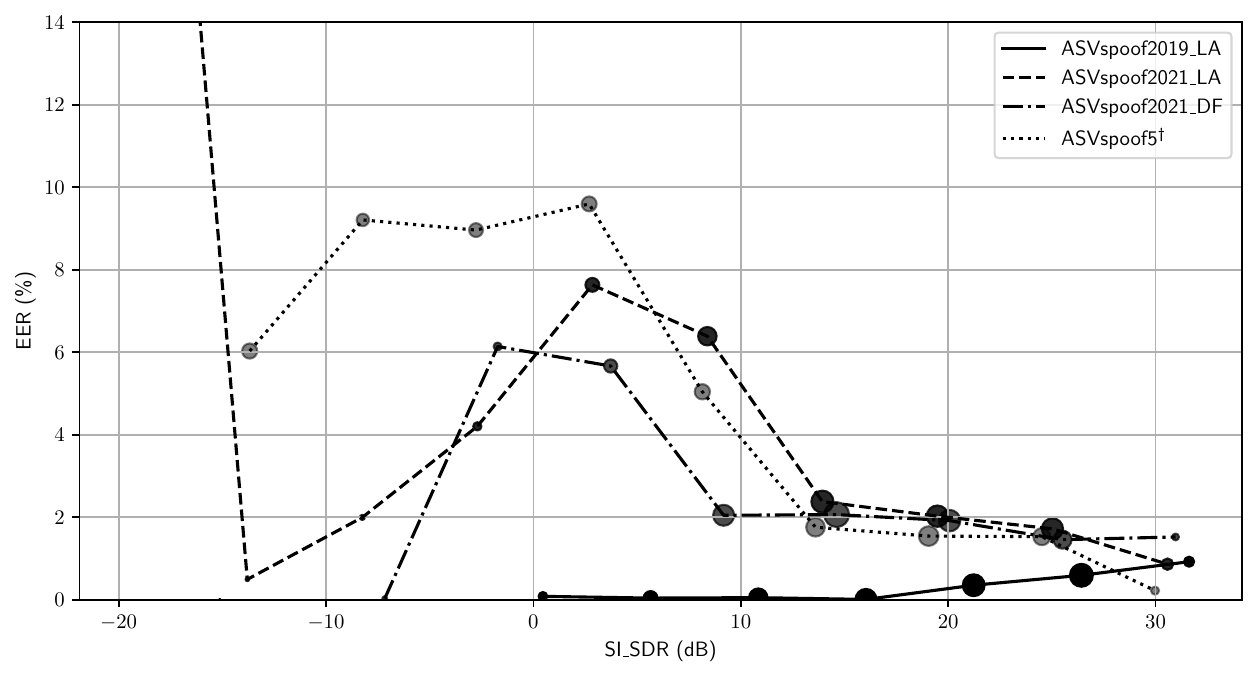}}
\caption{EER Distribution Across SI-SDR Bins}
\label{fig:si_sdr_bins}
\end{figure}

This observation leads us to consider the broader factors that influence the reliability of fake detection scores, specifically the length and quality of the speech in the audio. Clearly, as an utterance becomes shorter, the task of identifying a sample as real or fake becomes increasingly difficult. Similarly, the noisier the sample, the more challenging it is to distinguish between the classes. Figures \ref{fig:duration_bins} and \ref{fig:si_sdr_bins} illustrate the relationship between duration and speech quality bins versus EER, respectively, and confirm our expectations. Speech quality was measured using the non-intrusive models provided in \cite{kumar_torchsquim_2023}, and the size of the markers in the figures is proportional to the bin counts. Outliers from low counts notwithstanding, the general trend is that EER improves with longer and cleaner audio. In Figure \ref{fig:si_sdr_bins}, the reader might notice that for ASVspoof 2019, the EER actually increases with SI-SDR. However, it is important to note that the EER is already very low, and we attribute this increase to normal variation in the data and in the speech quality predictions.

\section{Conclusions}\label{sec:conclusions}
Our research has confirmed the value of data augmentation in training deepfake detection systems, and we have introduced novel loss functions that were previously unexplored in the deepfake detection literature. Through our exploration, we have developed a state-of-the-art approach that significantly improves the generalization capabilities of audio deepfake detection systems, as evidenced by our empirical results.

However, it is important to recognize that our solution represents only one component of a comprehensive defense strategy against audio deepfakes. A more comprehensive approach must encompass a suite of tools designed to utilize the varying amounts of information available on a case-by-case basis. In particular, the amount of verified audio for a speaker often varies; for instance, a seasoned politician typically has much more verified audio available than the CEO of a start-up or a member of the general public. Therefore, a more complete solution must include tools that can leverage this prior, verified information. Tools such as robust speaker verification, scalable forensic methods (e.g., for detecting manual splicing), and models that capture the nuances of speaker cadence and volubility, as well as those for identifying a speaker's native language and conducting linguistic analysis, would be instrumental in making the best use of all available data.

As we continue to refine our detection methods, it is imperative that we also consider the broader implications of deepfake technology. By integrating our approach with a diverse set of analytical tools, we can forge a more robust and resilient defense against the constantly evolving threat of deepfakes.

\bibliography{references}
\bibliographystyle{IEEEtran}
\end{document}